\documentclass[preprint,pre,a4paper,superscriptaddress,showpacs,nofootinbib]{revtex4}

\usepackage{amsmath}
\usepackage{amsfonts}
\usepackage{amssymb}
\usepackage{bm}
\usepackage{latexsym}

\usepackage{graphicx}

\usepackage{float}

\newcommand{\rsfig}[1]{\begin{center}
                       \includegraphics*[width=0.8\textwidth]{{#1}}
                       \end{center}
                       }

\begin{document}
\sloppy

\title{ Plastic Response of a 2D Lennard-Jones amorphous solid:\\
Detailed analysis of the local rearrangements at very slow strain-rate.}

\author{A.~Tanguy}
\email[Email: ]{atanguy@lpmcn.univ-lyon1.fr}
\affiliation{
  Laboratoire de Physique de la Matière Condensée et Nanostructures
               Université Lyon 1; CNRS; UMR 5586
               Domaine Scientifique de la Doua
               F-69622 Villeurbanne cedex; France}

\author{F.~Leonforte}
\affiliation{
  Laboratoire de Physique de la Matière Condensée et Nanostructures
               Université Lyon 1; CNRS; UMR 5586
               Domaine Scientifique de la Doua
               F-69622 Villeurbanne cedex; France}
\author{J.-L.~Barrat}
\affiliation{
  Laboratoire de Physique de la Matière Condensée et Nanostructures
               Université Lyon 1; CNRS; UMR 5586
               Domaine Scientifique de la Doua
               F-69622 Villeurbanne cedex; France}
\begin{abstract}
We analyze in details the atomistic response of a model amorphous material submitted to plastic shear in the athermal, quasistatic limit. After a linear stress-strain behavior, the system undergoes a noisy plastic flow. We show that the plastic flow is spatially heterogeneous. Two kinds of plastic events occur in the system: quadrupolar localized rearrangements, and shear bands. The analysis of the individual motion of a particle shows also two regimes: a hyper-diffusive regime followed by a diffusive regime, even at zero temperature.

\end{abstract}


\maketitle

\section{Introduction}
\label{sec:intro}

During the last two decades, a large number of numerical  studies
 ~\cite{livre,tanguy02,tanguy02b,tanguy04,tanguy05,maloneylemaitre,maloney,Weaire,Maeda,Varnik,Rottler,falklanger,Malendro,Schuh,Nanda,Bailey,Stankovic,depablo,nagel,goldenberg2002,cavaille,roux,vandembroucq,Krishna,picard}
has been devoted to  the peculiar mechanical response of amorphous
(disordered) materials. This keen interest can be related to the
recent synthesis of metallic amorphous glasses~\cite{Synthese}
whose mechanical properties  compare very favorably  to those of
crystals with similar compositions
\cite{livre,Synthese,Met1,Met2,Met3,Met4}. More generally, many
``soft'' materials  (foams~\cite{Argon,kabla,debregeas,Dennin,Kraynik,FoamLiu,FoamLiu2,cantat}, granular packings~\cite{grains,HJH1,HJH2,HJH3}, 
pastes~\cite{coussot,coussotshear,colloidal}), are also
characterized by a disordered microscopic structure, and their
mechanical and rheological properties are associated with
deformations of this structure. They can be studied either as
mesoscopic analogs of ``hard'' glasses, or in many cases for the
intrinsic interest of their rheological properties.

Both the elastic and plastic response of  amorphous materials are
currently studied, numerically and experimentally. In the elastic
limit,  numerical investigations on model materials (Lennard-Jones
glasses) have shown the essential role of a non-affine
contribution to the displacement field. This contribution, which
vanishes in a standard, homogeneous elastic material,  is
organized  in  rotational structures on  a mesoscopic scale
~\cite{tanguy02,tanguy02b,tanguy04,tanguy05,maloneylemaitre,maloney,Weaire}.
These rotational structures store a substantial part of the
elastic energy, and are responsible  for the low value of the
measured shear modulus ~\cite{tanguy02b,maloneylemaitre,Weaire}.

On the onset of plastic, irreversible deformation, important
changes are observed in the non-affine displacement field. In
contrast to the case of plasticity in crystals, the microscopic
description of plasticity in amorphous materials is still
incomplete and controversial
~\cite{livre,Rottler,falklanger,Chaudhari}.  Some points are now
well established.  When the behavior becomes irreversible, that is
after a ``spinodal'' (in the mechanical sense)  limit has been
reached by the system ~\cite{Malendro}, the deviation from the
affine response becomes localized ~\cite{tanguy05,maloney,Maeda}.
This deviation was identified first by Argon ~\cite{Argon} and
described by Falk et al. ~\cite{falklanger} in terms of ``Shear
Transformation Zones" (STZ). These ``STZ" have been considered for
a long time as elementary processes for a mean-field treatment of
the visco-plastic behavior of foams and emulsions. In the case of
foams, these local rearrangements are associated with T1 events
~\cite{Argon,kabla,debregeas,Dennin,Kraynik,FoamLiu,FoamLiu2},
while a precise identification of these rearrangements in
molecular glasses (like the metallic glasses for example) is still
a matter of debate ~\cite{Met3}. Experimentally, at a larger
scale, the plastic flow of these disordered dense systems gives rise to
a heterogeneous flow behavior, where a large shear band coexists
often with a frozen region
~\cite{coussot,coussotshear,colloidal,grains,Met1,Met2}.
Theoretically, it has been shown, using mesoscopic numerical
models ~\cite{vandembroucq,Krishna,picard} with long range
elasticity, that individual plastic events (like the STZ) can
concentrate statistically to create large scale fragile zones
where the deformation of the system takes place. However, a
multiscale description including a realistic description of the
local plastic events is still lacking, and the present mesoscopic
models are based on empirical assumptions for the occurrence and
the shape of the local plastic rearrangements.

In this paper, we study in details the local plastic
rearrangements occurring in a model Lennard-Jones glass submitted
to a quasi-static shear, at zero-temperature. 
We first identify the elementary plastic
processes, and there spatio-temporal statistical correlations.
This part of our work is closely related to a recent study by
Lema\^{i}tre and Maloney~\cite{maloneylemaitre,maloney}, and our results are consistent with their
observations, while not described in the same way. 
We then study the diffusive behavior of an individual particle 
immersed in this plastic flow.

\section{Sample preparation and quasi-static deformation procedure}
\label{technical}

The systems we study are slightly polydisperse two-dimensional
Lennard-Jones glasses, described in detail in
Ref.~\cite{tanguy02b}. Initial configurations at zero temperature
are obtained using the quenching procedure described in
Ref.~\cite{tanguy02b}. A    liquid of $N=10 000$ spherical
particles interacting via simple Lennard-Jones pair potentials
(characterized by an interaction diameter $\sigma$) is quenched,
using a conjugate gradient method, into the nearest energy
minimum. The average density and the starting temperature are, in
Lennard-Jones units, $\rho=0.925\sigma^{-2}$ and $T*=2$.

After the quench, two layers of particles, with thickness
$2\sigma$, are singled out and assumed to constitute parallel
solid ``walls'' that will impose the deformation to the system. The
resulting ``shear cell'' has a thickness $L_y=100\sigma$  (distance
between walls) and a width $L_x=104\sigma$. Other system sizes
(but with the same density $\rho$) have also been checked when
necessary: namely $(L_x,L_y)=(20,196),(30,296),(40,396),(50,496)$.
The configurations are then submitted to a quasi-static imposed
shear, by applying constant displacement steps  $\delta u_x$ to
the particles of the upper wall, parallel to this wall, and
keeping the lower wall fixed. After the displacement $\delta u_x$
has been imposed to the upper wall, the entire system is relaxed,
with fixed walls, into its new closest equilibrium position. The
equilibrium position is defined here as a minimum in the total
potential energy, hence the evolution of the system is studied at
zero temperature. Strictly speaking,  an ``athermal, quasi-static''
deformation \cite{maloneylemaitre} corresponds to the procedure
described above, in the limit $\delta u_x \rightarrow 0$.

It is important to understand the physical situation associated
with such a ``quasi-static'' procedure. If we consider a glassy
material (soft or hard) at finite temperature, we can define two
different characteristic times. The first one, $\tau_{diss}$, is
the time it takes for a localized energy input to spread over the
whole system and be dissipated as heat. The corresponding
mechanisms can be viscous (in a soft material) or associated with
phonon propagation in a metallic glass. The quasi-static procedure
corresponds to a shear rate, $\dot{\gamma}$, much smaller than
$\tau_{diss}^{-1}$.  A second, much longer time is the structural
relaxation time of the system, $\tau_{relax}$, associated with
spontaneous aging processes that take place within the system in
the absence of any external drive. By quenching after every
displacement step, we prevent any such processes from taking
place, meaning that the equivalent shear rate is larger than
$\tau_{relax}^{-1}$. In this simplified picture, the plastic
response of  a glassy system driven at a shear rate smaller than
the inverse relaxation time, which corresponds to many
experimental situations both for hard and soft systems, should be
reasonably well described by the quasi-static approach. This
picture, however, is oversimplified. The relaxation of the system
is in general stretched, meaning that relaxation processes take
place over a broad spectrum of times. The quasi-static approach
ignore the ``fast'' wing of this relaxation spectrum, which would
take place in real experiments at a finite value of
$\dot{\gamma}$. Nevertheless, all these relaxation times are very large 
at small temperature. 

In our case, the ``quasi-static'' procedure depends more significantly 
on the choice of the finite elementary displacement step. 
The next issue is thus to determine the value of the elementary
displacement step, $\delta u_x$, that can be considered to be a
reasonable approximation of the limit $\delta u_x \rightarrow 0$.
>From our previous simulations ~\cite{tanguy02b,tanguy04,tanguy05}, we know
that for systems of size $L=100\sigma$, prepared with the same quench protocol, the elastic character of
the response is preserved in average for shear strains smaller than
$10^{-4}$. By choosing an elementary  displacement of the wall
$\delta u_x = 10^{-2}\sigma$, we obtain an elementary  strain step
$\delta\epsilon_{xy}\equiv \delta u_x/(2.L_y)\approx 5. 10^{-5}$,
clearly below this threshold. In order to allow the system relax between 
two successive elementary steps, the time ellapsed must be $\Delta t\gg \tau_{diss}$, and the corresponding shear rate is thus limited by the additional condition $\dot\gamma \ll \delta\epsilon_{xy}/\tau_{diss}$. 

 Finally, we note that the quasi-static procedure ensures that,
 after a local ``bond'' breaks, new bonds can form instantaneously.
 In a simple van der Waals system such as the one under study,
 this makes it very unlikely that the material could fracture, at
 least under the type of volume preserving deformation we are studying.

In the following, we will discuss the numerical results obtained
for the onset of plastic deformation by averaging over 20
configurations. Each of these configurations has been subjected to
5000 elementary deformation steps, corresponding to a total strain
of 25\%. In order to study the ``stationary'' regime obtained for
large plastic deformation, we will also discuss results obtained
for a system deformed during 34000 steps, corresponding to a
total shear strain of 170\%.

\section{Stress-Strain relation: elastic thresholds and plastic flow.}
\label{Part1}

\begin{figure}[t]
\rsfig{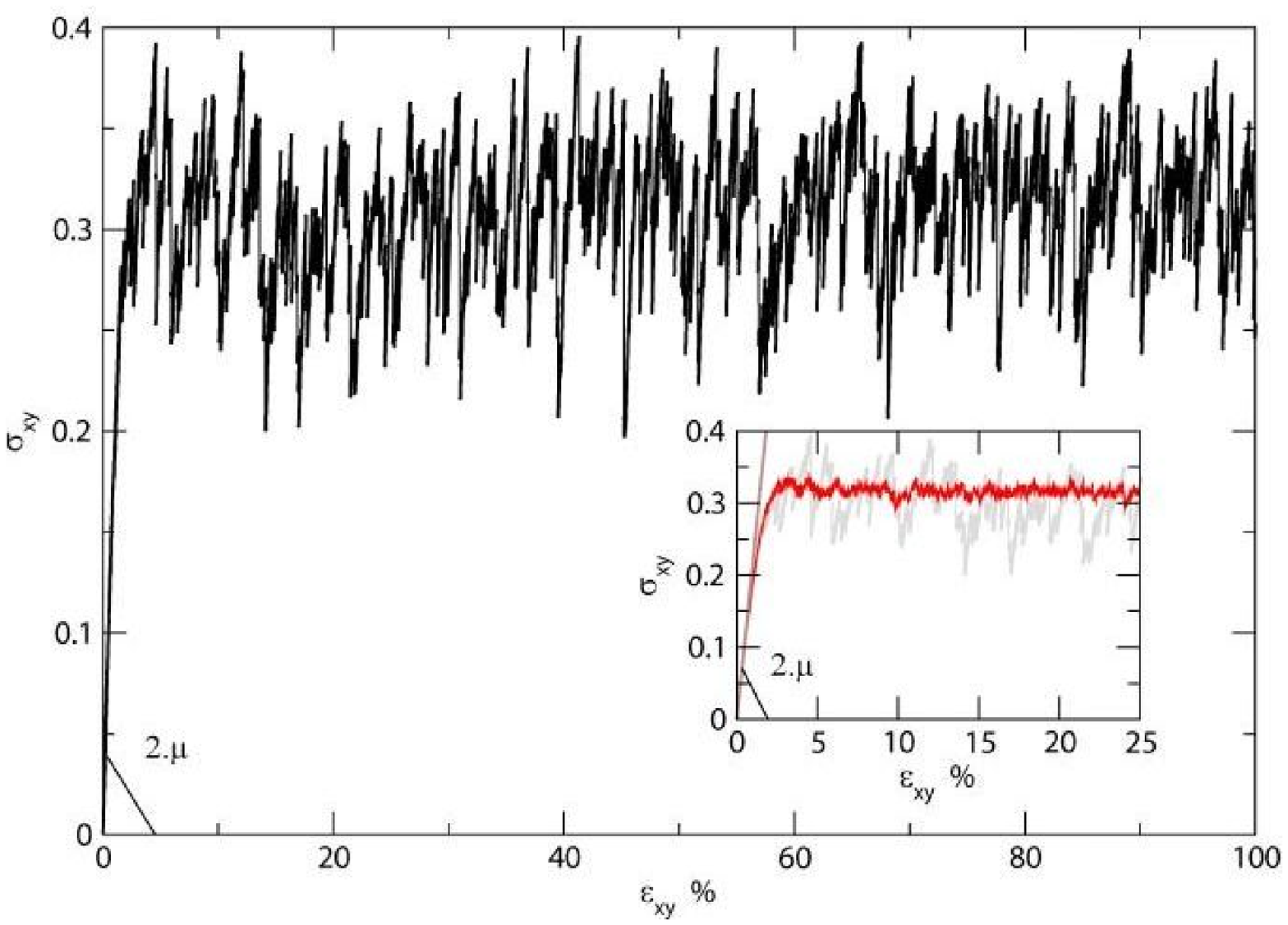}
\vspace*{0.8cm}
\caption[]{ (Color online)
Averaged shear stress $\sigma_{xy}$ as a function of the shear strain $\epsilon_{xy}$ applied at the borders. Two different regimes are shown: a linear regime and a regime of plastic flow. Note that the elastic (reversible) part of the linear regime is restricted to strains up to $10^{-4}$. The rest of the linear regime is noisy, with irrevesible plastic rearrangements.
Inset: Same stress-strain relation (grey) and average over 20 configurations. The plastic flow displays a plateau.
\label{figcomport}}
\end{figure}

We start by discussing the ``macroscopic'' (i.e. computed for the
whole sample) stress-strain relation. We show in
Fig.\ref{figcomport} the shear stress $\sigma_{xy}$  as a function
of the total shear strain $\epsilon_{xy}$, computed from the
boundary displacement. The stress is obtained from the usual
microscopic Irving-Kirkwood definition~\cite{tanguy02b,Kirkwood}, 
and averaged over the whole system. Results
are shown for a single realization, and for an average over 20
realizations.  From this figure, we identify two different
regimes.  A linear   increase of the stress as a function of
strain, followed by a very noisy plastic flow, with constant
average shear stress.

Although it appears to be elastic in the usual, averaged sense,
the  linear part of the stress-strain curve  is not strictly
reversible at the microscopic level. In fact, the elastic
reversible part is restricted for our systems to $\epsilon_{xy}
\le 10^{-4}$, as already shown in a previous paper
~\cite{tanguy02b}. For $10^{-4} \le \epsilon_{xy} \le
\epsilon^{(p)}_{xy}$, the shear stress evolves mainly linearly
with increasing strain, but displays small jumps, giving evidence
of irreversible energy losses. Then the stress-strain curve
saturates. The upper threshold ($\epsilon^{(p)}_{xy}\approx
0.015$) does not seem to depend on the system size at least for
the configurations we have checked. Its value is not far from the
one deduced from experiments  in metallic glasses~\cite{Met2}. It
corresponds approximately in our case to a displacement of one
particle diameter $\sigma$ at the border of a vortex-like
structure of size $\xi\approx 30\sigma$ as shown in the elastic
inhomogeneous response of these systems
\cite{tanguy02,tanguy02b,tanguy04,tanguy05}
($\epsilon_{xy}^{(p)}\approx \sigma/2.\xi = 0.017 $). After the
linear increase, the stress thus reaches a very noisy plateau,
characteristic of plastic flow. In this second part of the
stress-strain relation, the fluctuations of the measured shear
stress are enhanced (see figure \ref{figdrops}).

The distribution of the stress jumps is shown in figure
\ref{figdrops}. We show in this figure that the stress jumps in
the linear part (straight line in the figure) are much smaller
than the stress jumps in the noisy ``plateau'' region. The average
energy loss (that can be obtained as the average stress jump times
the elementary imposed deformation) is thus by at least one order
of magnitude larger in the plateau than in the previous part. This
suggests rather different types of microscopic response in the two
different regions. Note that the distribution of the stress jumps
is not strictly exponential as in Ref.~\cite{maloney}, but shows a
marked deviation from the exponential decay that could be due to
the difference of boundary conditions. Indeed, it is known that
the presence of walls affects strongly the plastic behavior of
glasses (see for example Ref. \cite{Varnik}). However, the size
dependence of the stress jumps (figure \ref{figdrops}-b)
 seems to be in agreement with the
scaling $\Delta\sigma_{xy}\propto1/L_{y}$ already proposed in
Ref.~\cite{maloney} and related to the existence of a characteristic
displacement $\approx\sigma$ resulting from the instability. 
Note finally that the situation is
quite different from the one observed in crystalline plasticity
\cite{brechet}, in which the characteristic size of stress drops
in the linear and in the plateau region is rather similar, although the
shape of the distribution may differ in both parts.

Before we start analyzing  the response of the system at the
atomic scale in these two different regimes, it is interesting to
look at the cumulative displacement in the direction of the
imposed shear. This is shown  in figure \ref{figshearbande}, where
the cumulative displacement of the particles has been averaged for
each distance $y$ from the shearing wall on a layer with a
thickness of few particle diameters (typically $4.0 \sigma$), for
a total deformation here of 25\%. In general, the cumulative shear
appears to be very heterogeneous, even for such large strains. In
some cases (like in figure \ref{figshearbande}) a broad shear band 
appears in the center of the sample, while
the particles close to the boundaries are dragged by the wall
motion. In other cases several parallel sheared regions coexist,
separated by unstrained regions. Note that, in all our samples, the
cumulative shear zones are observed  to occur away from the
boundaries, unlike in experiments and simulations on foams in a 2D
Couette Geometry ~\cite{kabla,debregeas}, or on concentrated
emulsions ~\cite{coussotshear}, where the shear is localized at a
boundary, but closer to experiments on metallic
glasses~\cite{Met1,Met2,Met3,Met4}, or on granular systems
~\cite{grains,HJH1,HJH2,HJH3}. This is one of the marked
differences appearing in different amorphous systems.

We now turn to the detailed analysis of the atomistic response of the system.

\section{Detailed analysis of the local plastic rearrangements}
\label{Part2}

\begin{figure}[t]
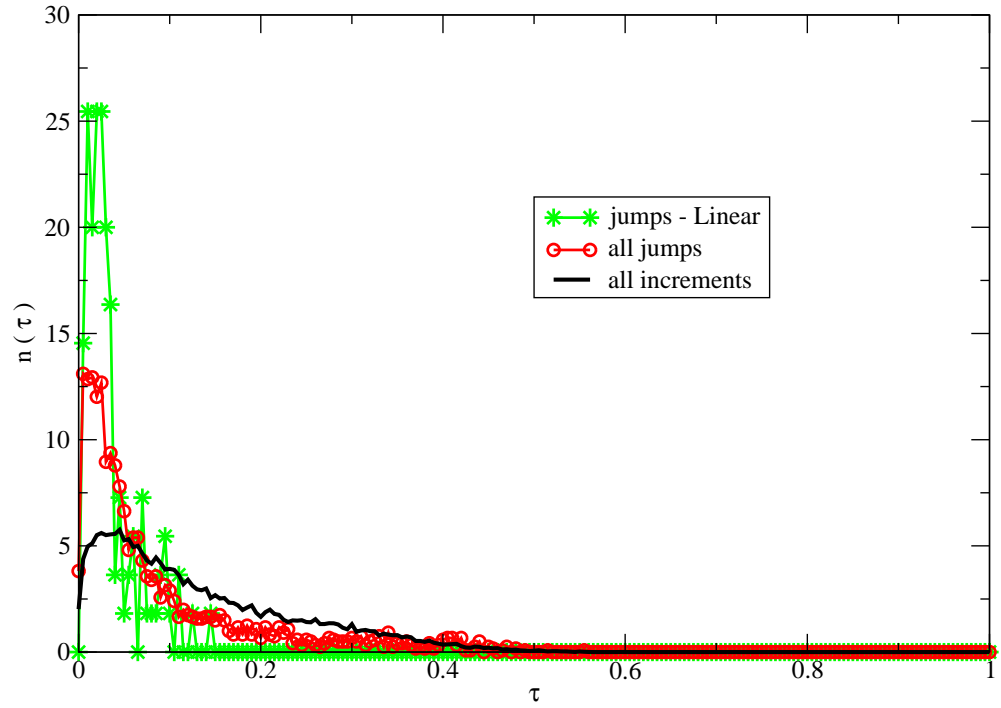

\rsfig{FIG-PLAST/FIG-ART/histTP-Jumps.eps}
\vspace*{0.8cm}
\caption[]{(Color online)
Normalized distribution of the participation ratios $\tau$ of the non-affine field, for all steps (line), for all events associated to the stress drops (circles), and for the events associated with a stress drop but only in the linear regime (stars).
\label{figtpart}}
\end{figure}

In the previous section, we inferred from the distribution of
stress drops in the linear and plateau regions, respectively, that
 different microscopic events were taking place in these two
 regions.
 Following the distinction made previously between the
dissipative behavior in the linear regime, and the one in the
regime of plastic flow, we can distinguish two kinds of plastic
events. In the following, we will distinguish between these events
based on the participation ratio  $\tau$ for the  non-affine
displacement field associated with the event. $\tau$ is defined as
$$\tau\equiv\frac{1}{N}.\frac{\left(\sum_i\delta
u_{n.a.}(i)\right)^2}{\sum_i u_{n.a.}(i)^4}$$ where $u_{n.a.}(i)$
is the {\it non-affine} displacement of the particle $i$, that is
the displacement after its usual affine shear component
(corresponding to a uniform shear strain) has been removed.
Obviously,  $\tau\sim 1$ for events involving the whole system,
and $\tau\rightarrow 0$ in case of a localized event. From figure
\ref{figtpart} it appears that the linear part of the stress
strain curve is dominated by events with a much smaller value
of $\tau$ than the plateau region. Visual inspection allows a
clear distinction between two types of events associated with
small ( $< 0.05$) and large values of $\tau$, which we now describe
in more detail.

\subsection{Localized, quadrupolar rearrangements}

\begin{figure}[t]
\rsfig{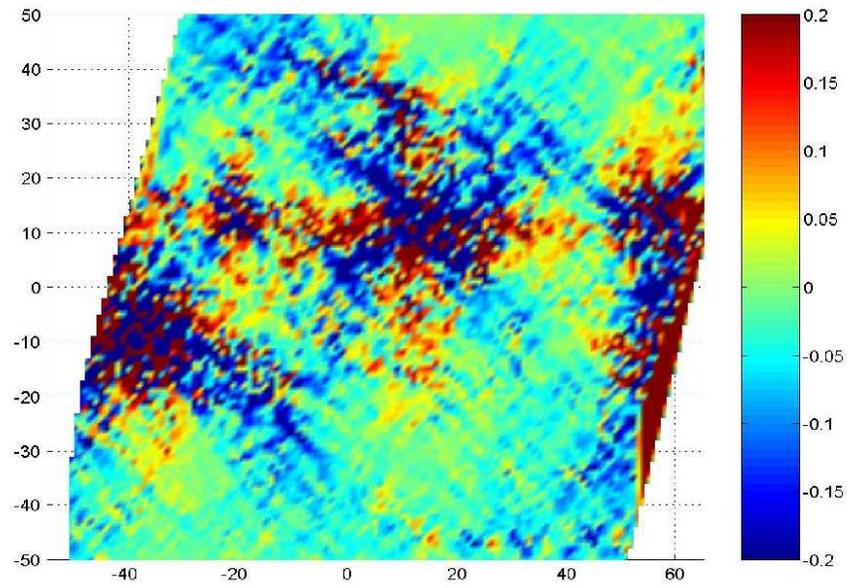} 
\rsfig{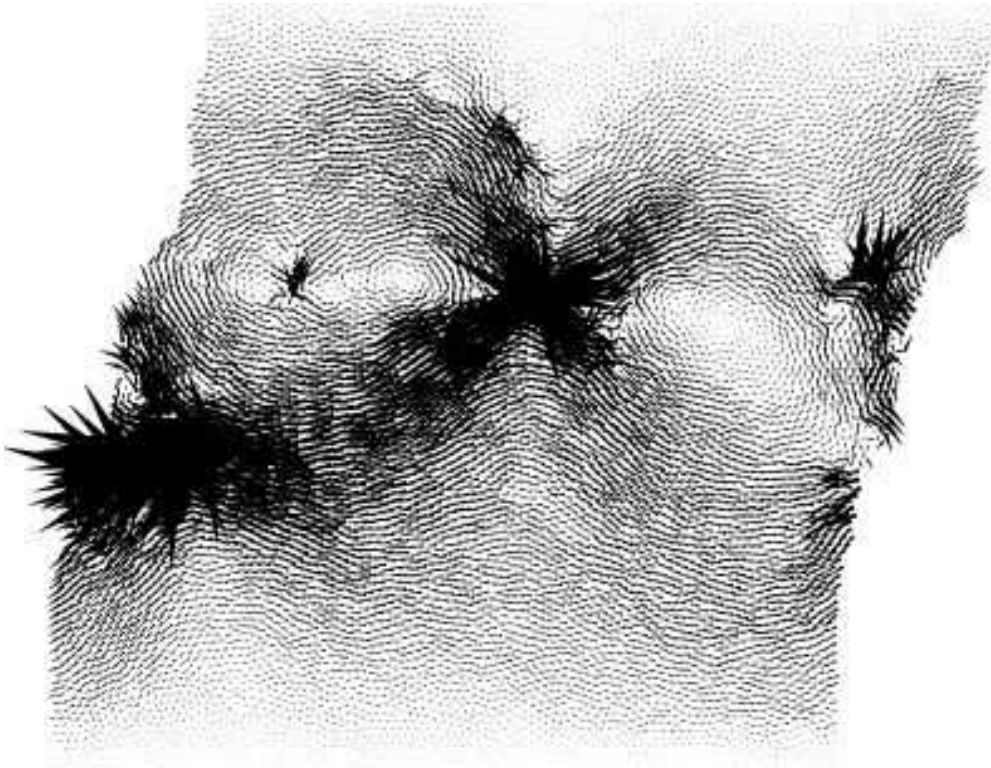} \caption[]{(Color online)
Changes in the local shear stresses during a localized plastic
event (left), and associated displacement field (right). In this
case, the plastic event involves a local rearrangement at the
border of two vortices. \label{figdWxyquadru}}
\end{figure}

A typical example of localized event is shown in  figure
\ref{figdWxyquadru}.  {\it All} dissipative events in the linear
part of the stress-strain curve are of this type. In contrast,
only very few stress drops in the ``flowing'' regime (plateau
region) are associated with such localized events. These events
give rise to the small stress drops in figure \ref{figdrops}, at
various yield stresses. They do not, however, contribute
substantially to  the horizontal displacement field shown in
figure \ref{figshearbande}.

Figure \ref{figdWxyquadru} displays  the non-affine displacement
of each particle in a typical localized event. The large
displacements indicate the location of the irreversible
deformation. The few particles involved in the rearrangement are
localized at the boundary between two adjacent vortices of the
non-affine field, that are reminiscent of the non-dissipative
(elastic) non-affine response of the system~\cite{tanguy02b}. It is possible to
identify the center of the plastic rearrangement by selecting  the
particle undergoing the largest non-affine displacement. This
particle is at the center of a redistribution of shear stresses
with an apparent symmetry characteristic of a quadrupole (figure
\ref{figdWxyquadru}). In this figure, the change in the local
shear stress is obtained from the usual Irving-Kirkwood
definition, as described in \cite{tanguy02b,Kirkwood}. In order to identify
more precisely as in Ref.~\cite{maloney} the symmetry of these rearrangements, we move to a
coordinate system $(r,\theta)$ centered on this point, and project
the corresponding radial and azimuthal part of the displacement
field onto circular harmonics $e^{i.n.\theta}$ (see figure
\ref{figproj}). For the $(n=2)$ (quadrupole) contribution, we get
the well known $1/r$ dependence of the radial projection,
corresponding to an homogeneous, linear and isotropic elastic
medium \cite{picard}. For unknown reasons however, the
$r-$dependence of the azimuthal projection is much more noisy in
this case. Note also that the $(n=3)$ contribution is far from
being negligible (not shown here), indicating that the
displacement field is not strictly quadrupolar, a deviation that
may originate from the boundary conditions we are using.

\begin{figure}[t]
\rsfig{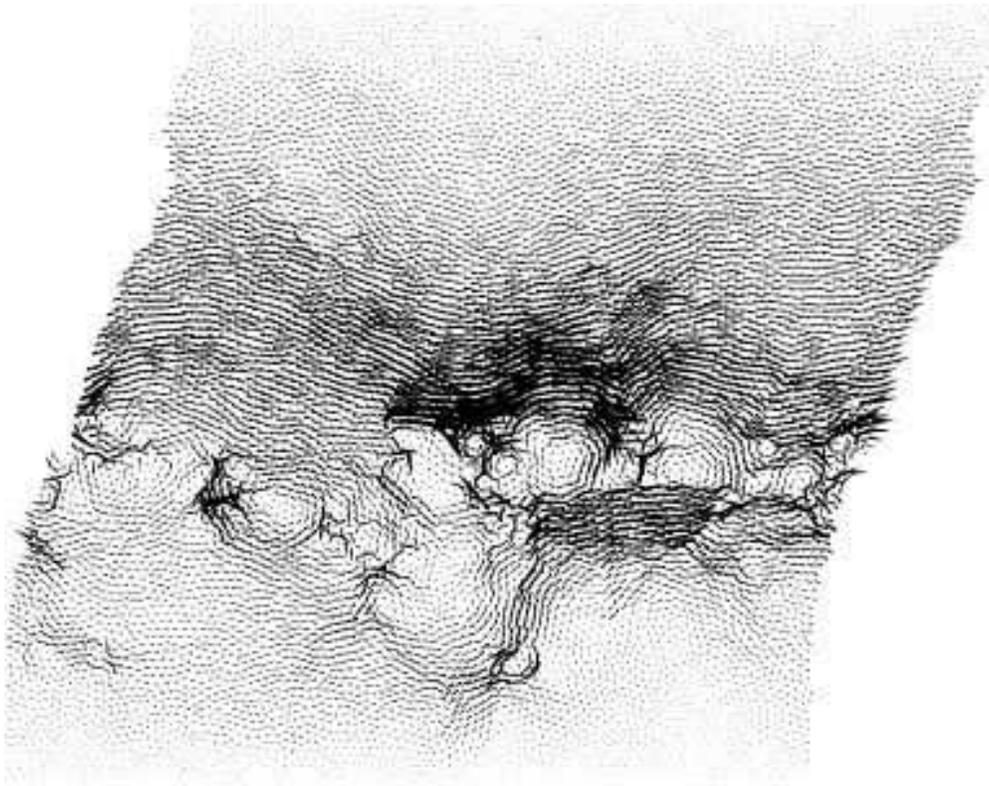}
\vspace*{0.8cm}
\caption[]{(Color online)
Displacement field associated to an elementary shear band. As it can be seen here, the shear band is due to an alignement of vortices, along the direction of shearing.
\label{figfield}}
\end{figure}

When the flowing  plateau is approached, many  such local
rearrangements tends to appear simultaneously  and to concentrate
spatially during a single plastic event. Eventually, when the
plastic flow regime is reached,  another kind of event appears,
involving a much larger amount of  dissipated energy (or stress
drop). These larger events are  made of an alignment of rotational
rearrangements (see figure \ref{figfield}) along the direction of
the imposed external shear. This second kind of event constitutes
the ``elementary shear bands", that we  now describe  in more
detail.

\subsection{Collective, large scale events: elementary shear bands}

\begin{figure}[t]
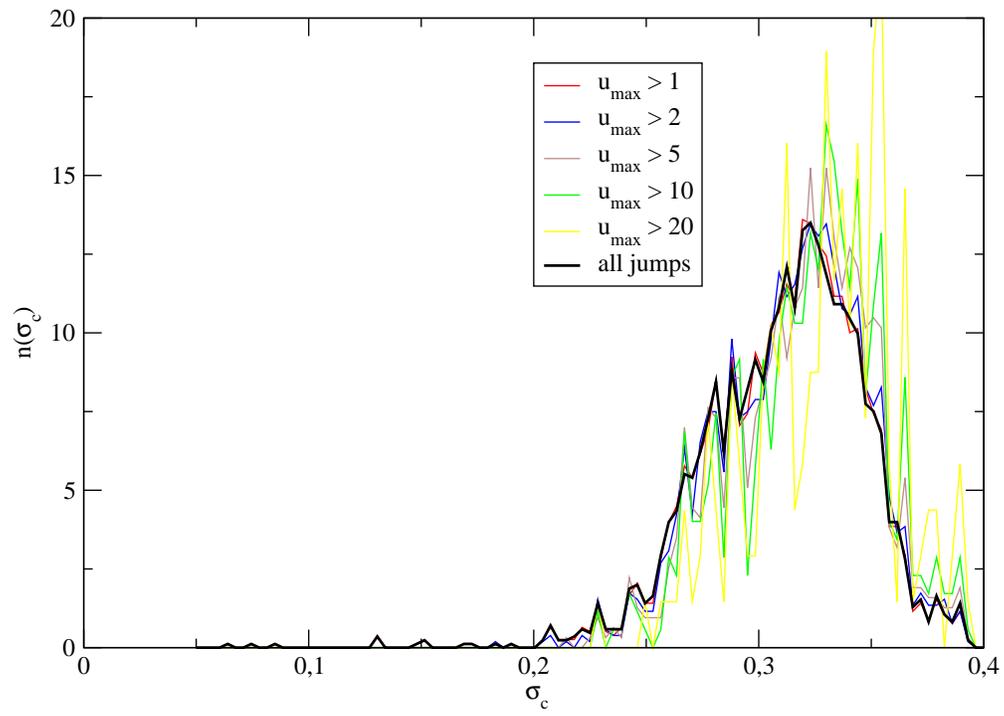

\rsfig{FIG-PLAST/FIG-ART/HistoSeuil100pts.eps}
\vspace*{0.8cm}
\caption[]{(Color online)
Distribution of plastic thresholds just before a plastic event occurs. This distribution does not depend significantly on the amplitude of the maximum $u_{max}$ of the non-affine displacement during the event
(contrary to the distribution of the energy drops).
\label{figthresh}}
\end{figure}

The collective events that dominate the flow behavior in the
plateau region involve a large displacement that spans the whole
length of the sample in the $x$ direction, and are localized in
the $y$ direction.
 It is easy to identify the center line of such an elementary
shear band, because the displacements of the particles are so high
that they lead to an inversion of the instantaneous ``velocity
field'' (i.e. the displacement within a single strain step) in the
direction of the sollicitation, above and below the elementary
shear band (figure \ref{figbande-ab}-a). The largest displacement
inside the sample (located at the edge of a shear band) can reach
more than 100 times the displacement imposed at the wall, i.e. it
is close to one particle size in our case. It is interesting to
note that the distribution of the global shear stresses at which
this kind of event occur (yield stress distribution), is unrelated
to the spatial  amplitude of the subsequent  event (figure
\ref{figthresh}). However, the largest events (that is with the
largest displacements) constitutes the enveloppe of the noisy
stress-strain relation (not shown here).

We see in figure \ref{figbande-ab}-b that these elementary shear
bands can take place anywhere in the sample, and not only at the
boundaries. In fact, the distribution of the distances between the
centers of successive shear bands (figure \ref{figbande-cd}-a) is
exponential, with a characteristic length $ \xi_B\approx 30\sigma$,
independent on the system size,
that corresponds to the size $\xi$ of the rotational structures
that have been identified  in the elastic response of the system
~\cite{tanguy02,tanguy02b,tanguy04,tanguy05}. The characteristic
distance between successive elementary shear bands in our system
is thus equal to the width of the elementary shear band itself.
Moreover, the Fourier transform (not shown here) of the temporal
evolution of the positions of the centers of the bands (figure
\ref{figbande-ab}-b) shows a $1/f^{0.5}$ behavior characteristic of a
random, sub-diffusive signal. All these results mean that the elementary
shear bands propagate essentially in a random walk manner, with a
step size of approximately $30\sigma$, confined by the two
boundaries.

The distribution of  distances covered by the upper wall between
successive occurrence of elementary shear bands (figure
\ref{figbande-cd}-b) is also exponential. It shows a
characteristic length $l_c$ ($l_c\approx 0.13$ in our case),
equivalent to a characteristic number of quasistatic steps (here
$13$ steps of amplitude $\delta u_x=10^{-2}$ on the upper wall).
This length does not depend significantly on system size. We can
find an explanation for the order of magnitude of this length
$l_c$, by dividing it by the radius $\xi/2$ of a vortex. If we
assume that the deformation is localized within a ``weak'' region of
thickness $\xi/2$, and that the system outside this shear band is
essentially unstrained,  the characteristic distance $l_c$ covered
by the wall between successive occurrence of elementary shear
bands corresponds to a deformation of $2l_c/\xi \simeq 1\%$ within
the weak region. This order of magnitude is approximately the one
 that corresponds  to the elastic threshold
$\epsilon^{(p)}_{xy}$ for the strain within the elementary shear
band. The elementary shear bands can thus be seen as weak
locations  where all the deformation concentrates, giving rise -
from a given local strain threshold - to a large plastic event
that relaxes all the accumulated elastic energy. The next shear
band event is spatially strongly correlated, within a distance
$\xi$.

Within this picture, we can simply describe the construction of
the plastic flow ``velocity'' profile as a diffusive process. For a
sample with transverse size $L_y$, the number of bands that are
created by a total strain $\epsilon$ is $\epsilon L_y/l_c =
(\epsilon L_y) /(\epsilon^{(p)}_{xy}\xi)$. If the bands are
created in a spatially correlated manner, with typical distance $\xi$, this
will result in an effective diffusion coefficient for these
plastic events of the form $D_{eff} = \xi^2 \times ( L_y
/\epsilon^{(p)}_{xy}\xi)$ (here the strain $\epsilon$ plays the
role of time). The shear will diffuse through the sample over a
``time'' scale $L_y \epsilon^{(p)}_{xy}/\xi$.
For our samples ($L_y=100$, $\xi=30$) the corresponding strain is
small, and the boundaries will almost immediately limit shear band
diffusion. The shear profile then is created by essentially
independent bands. An essentially homogeneous profile will be
obtained when the shear band density becomes of the order of the
inverse of the particle size, i.e. $\epsilon \sim
\epsilon^{(p)}_{xy} \xi/\sigma$.
In a  larger sample, on the other hand, this picture suggests that
the time (or strain) scale for establishing an homogeneous profile
may be very large, in fact proportional to system size, which
could explain the commonly observed tendency towards shear
localization in such systems.

In the next section, we explore the effect of the two kinds of
dissipative events described here (quadrupolar events and
elementary shear band), on the local dynamics of the particles.

\section{Diffusive trajectory of an individual particle.}
\label{Part3}

\begin{figure}[t]
\rsfig{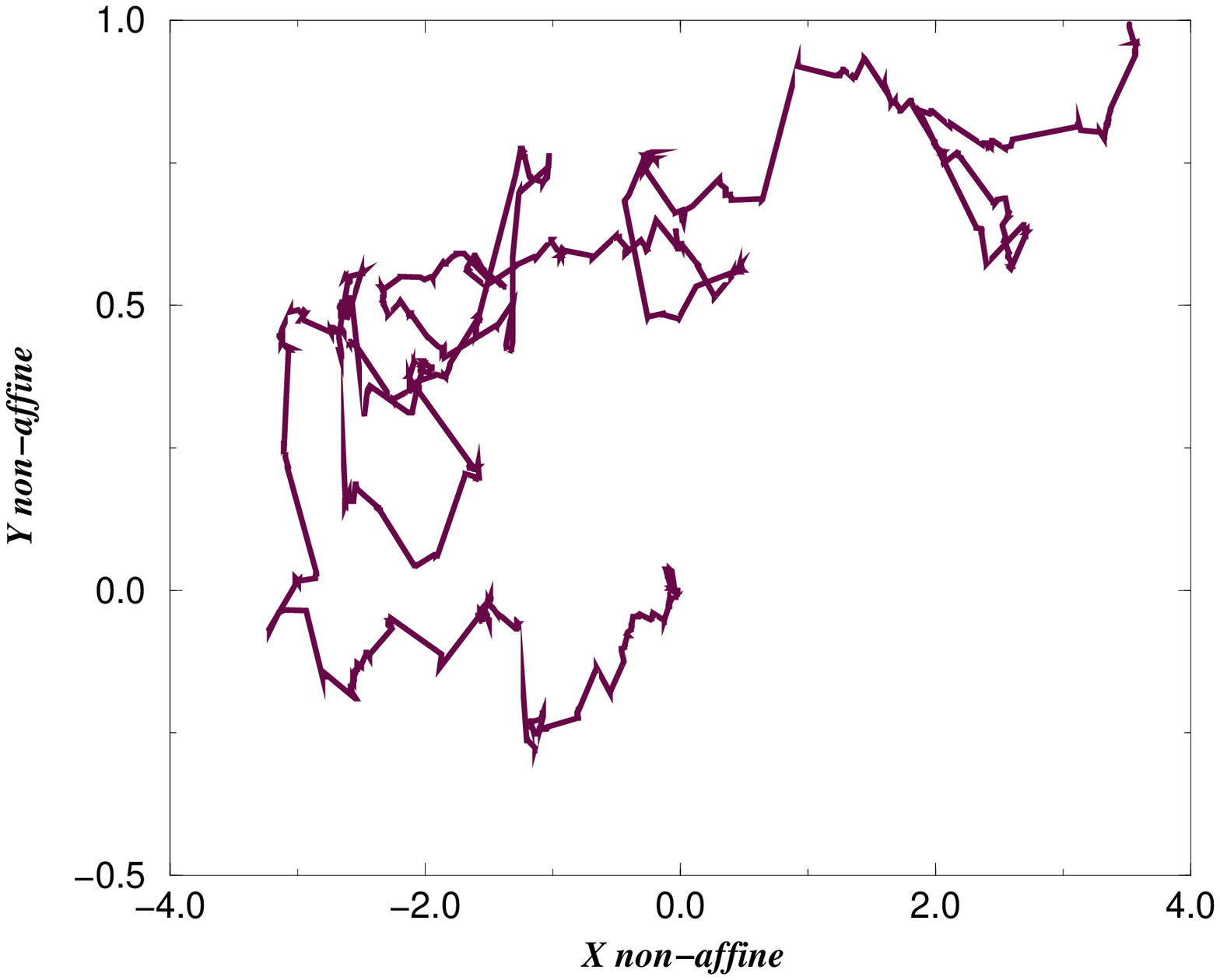}
\rsfig{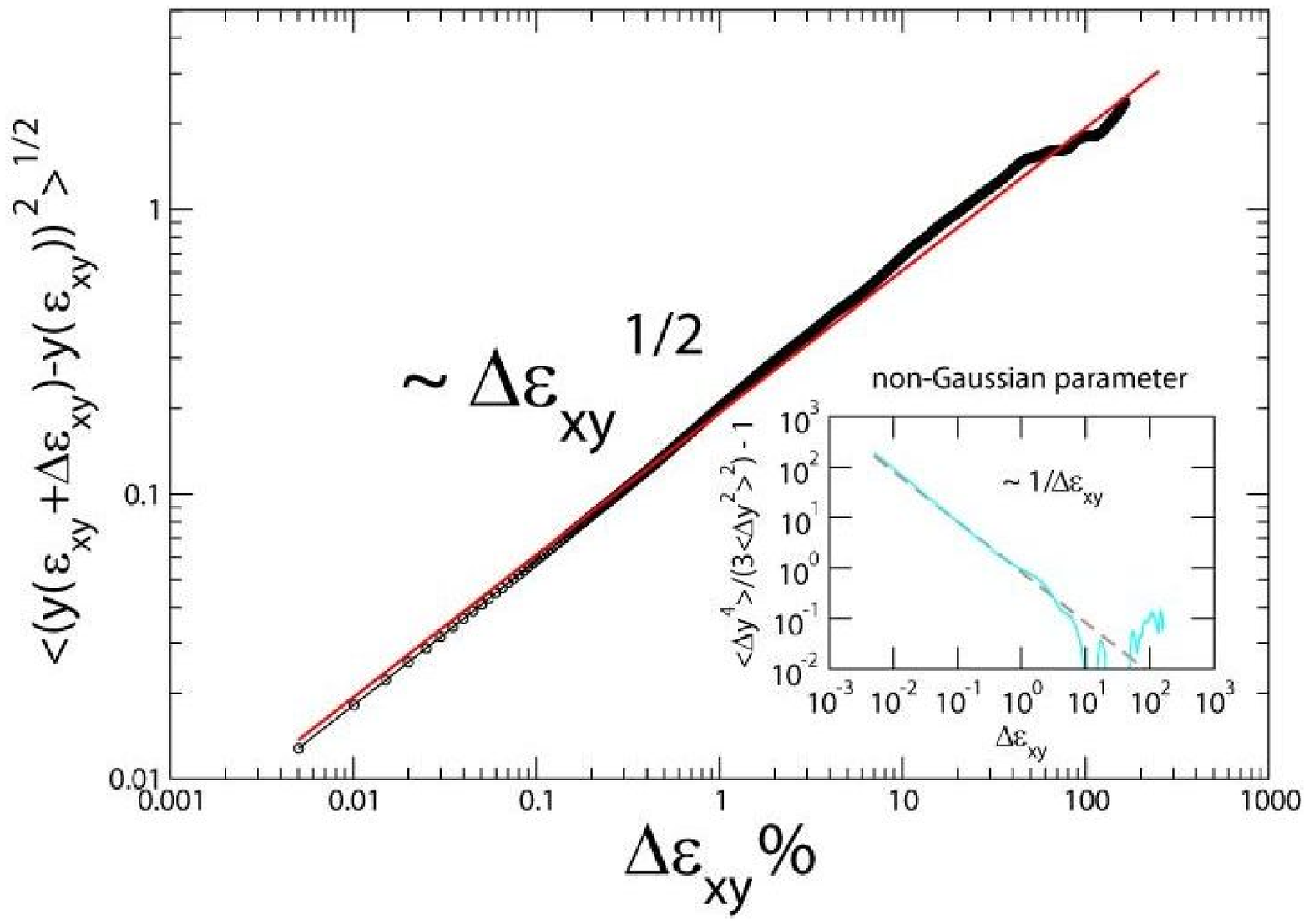}
\vspace*{0.8cm}
\caption[]{(Color online)
(a) Diffusive trajectory of one particle with respect to the affine displacement expressed in the units of mean particle diamater $\sigma$ for a cumulative shear strain of $25\%$. The displacement may be compared with the approximate cell dimensions $L_x = 104 \sigma$ and $L_y= 100 \sigma$.

(b) Variance $\left<\left(\Delta y(\epsilon_{xy}+\Delta\epsilon_{xy})-\Delta y(\epsilon_{xy})\right)^2\right>^{1/2}$ of the transverse coordinate of the individual particles, with the initial position $y(0)=10\pm 4$, as a function of the incremental strain $\Delta_{xy}$. Inset: non-Gaussian parameter of the transverse coordinate $\Delta y(\epsilon_{xy})$. The non-Gaussian parameter is far from being negligeable at small imposed strain $\Delta\epsilon_{xy}$.
\label{figsuivi}}
\end{figure}

Due to the average shear flow, the motion of each individual
particle is highly anisotropic. The motion in the direction of the
sollicitation will mainly indicate the presence of a central
sheared zone, while the motion in the transverse direction is zero
on average. A diffusive contribution to the motion of an
individual particle can be defined by removing the convective part
of the motion (affine displacement) in the $x$ direction. A
typical example of the resulting motion is shown in
figure~\ref{figsuivi}-a. In the following, in order to avoid
potential ambiguities associated with the inhomogeneous character
of the convective displacement along the shear direction, we
concentrate on a statistical analysis of the motion in the
transverse, $y$ direction.

\begin{figure}[t]
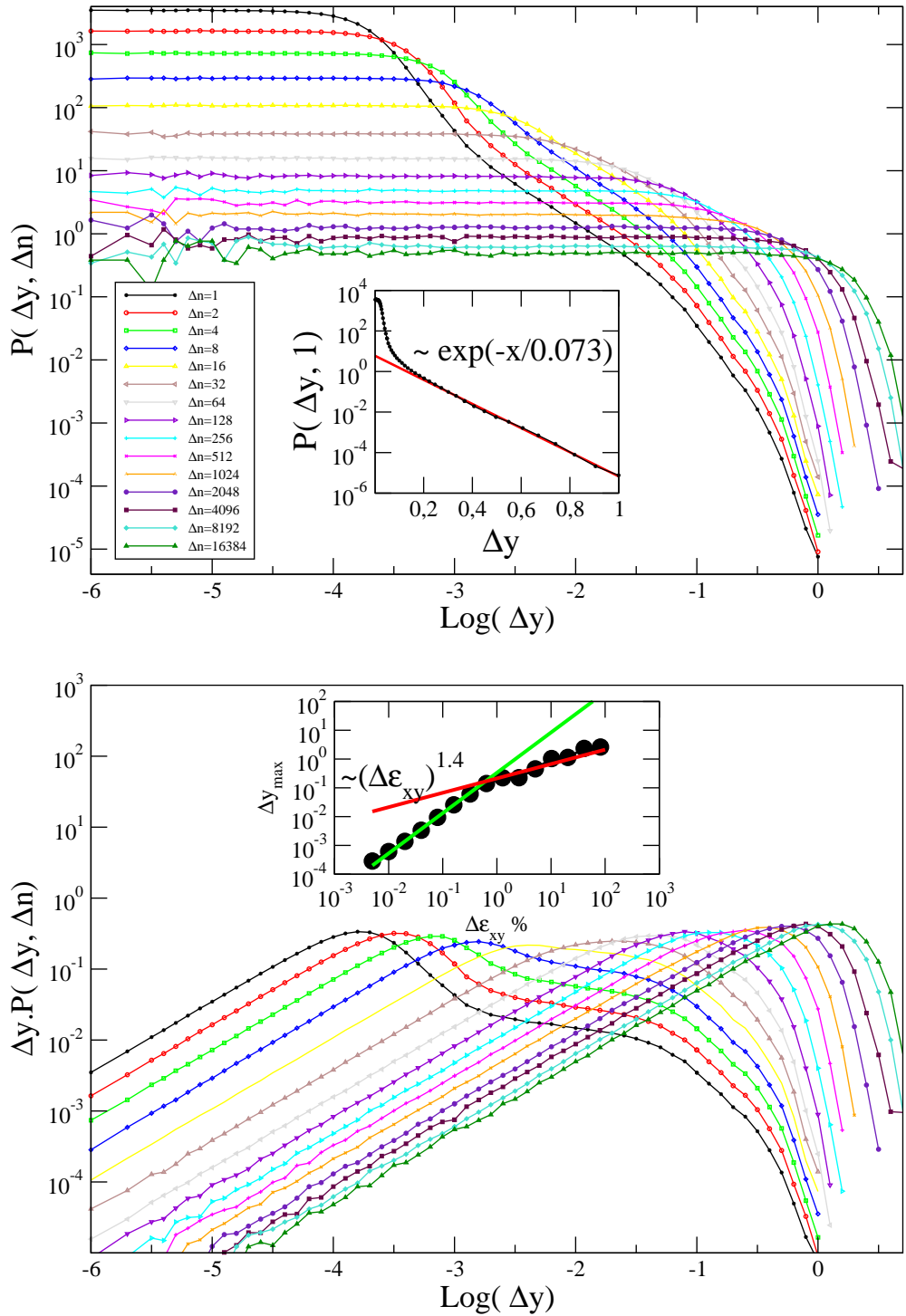

\rsfig{FIG-PLAST/FIG-ART/FuruLog2.eps}
\rsfig{FIG-PLAST/FIG-ART/FuruLogY2Eps.eps}
\vspace*{0.8cm}
\caption[]{(Color online)
(a) Distribution $P(\Delta y,\Delta n)$ of distances $\Delta y$ between the y positions of a particle of initial position $y(0)=10\pm4$, after a certain number $\Delta n$ of incremental strains has been imposed at a wall. The corresponding global strain is thus $\Delta\epsilon_{xy} = \Delta n.\delta u_x/2.L_y = 5.10^{-5}.\Delta n$.
Inset: Log-Linear plot of the distribution of elementary jumps of the particles in the y-direction. This distribution is a power law at small distances, with an exponential cut.

(b) Same distribution of distances $\Delta y$, but multiplied by the distance $\Delta y$ itself. The position $\Delta y_{max}$ of the peak appearing at small $\Delta\epsilon_{xy}$ (small $\Delta n$) evolves like $(\Delta \epsilon_{xy})^{1.4}$ (hyper-diffusive behaviour), while the peak appearing at large $\Delta\epsilon_{xy}$ evolves diffusively (like $(\Delta \epsilon_{xy})^{0.5}$). See inset.
\label{figfuru}}
\end{figure}

The random motion of a particle can be described by the
distribution of the size of its elementary jumps, and by the
temporal correlations between jumps~\cite{Crank}.  We find (see
figure~\ref{figfuru}-a for $\Delta n=1$) that the distribution
of the size of the transverse components of the elementary jumps
is symmetric, with zero average and finite variance $<\delta
y^2>^{1/2}\approx 10^{-2}.\sigma$. This finite variance results
from an upper exponential cutoff. It implies that, in the hypothetic 
absence of temporal correlations between jumps, the motion of
the particle should be diffusive at large enough times.

This seems to be confirmed by the study of the mean squared displacement
$<\Delta y^2> = <(y(\epsilon)-y(0))^2>$ of the particles in the
transverse direction, which grows essentially linearly with the
strain $\epsilon$ (see figure~\ref{figsuivi}-b), and allows one to define a 
diffusion coefficient. But although the motion seems to be  
diffusive by looking only at the second moment of the distribution, 
the study of the non gaussian parameter
$<\Delta y^4> /3 <\Delta y^2>^2 -1$ (inset in figure~\ref{figsuivi}-b)
shows that the situation is much more complex, 
particularly at short times, with a
markedly non-gaussian distribution for displacements smaller than
typically one particle size.

This deviation from gaussianity can be explored further through
the  distribution $P(\Delta y, \epsilon)$ of the transverse
distances $\Delta y$ between the positions of a particle, after a
total strain $\epsilon$  has been imposed to the sample
~\cite{Frac}. The distribution $P(\Delta y, \epsilon)$ is shown in
figure \ref{figfuru}-a. It can be seen as a quasistatic
equivalent of the van Hove distribution correlation function,
which is a standard tool to characterize diffusion in glasses and
supercooled liquids \cite{Diff1}. For a given $\epsilon$, the
function starts with a plateau, followed by a power-law decay, and
ends with an exponential cutoff. The width of the power-law decay
depends on the total strain $\epsilon$. The function is clearly
very different from the Gaussian propagator of simple diffusion.

In fact, the beginning and the end of this power-law behavior are
not self-similar (i.e. cannot be rescaled in a form $f(\Delta y/
\epsilon^\beta)$).  For small $\epsilon$, the long distance
contribution evolves less rapidly  with $\epsilon$ than the short
distance part. This is also why the amplitude of the initial
plateau, $P(0,\epsilon)$, shows two different behaviors as a
function of $\epsilon$ (see figure~\ref{figfuru}-a): a rapid
decrease at small  $\epsilon$, followed by a slower decrease for
larger $\epsilon$ after the power-law decay has disappeared.

This behavior is also shown in figure \ref{figfuru}-b where
the function $\Delta y. P(\Delta y,\epsilon)$ is plotted as a
function of $\Delta y$. In this representation, the values of
$\Delta y$ that contribute most to the average displacement appear
as peaks.  It is clear that two main peaks are present. The first
one corresponds to very small displacements, and its positions
evolves as $\epsilon^{1.4}$. The second one, which corresponds to
the actual diffusive process, appears at distances of order
$\sigma$, and its position increases as $\epsilon^{0.5}$. As the
deformation is increases, the intensity shifts progressively from
the first peak to the second one. This result supports the idea of
two different relaxation mechanisms in amorphous glasses, even at
zero temperature ~\cite{Diff2,Diff3,Diff4,Diff5}.

To quantify the difference between the two regimes, we show in
figure \ref{figfuru}-b the position of the main peak in 
$\Delta y. P(\Delta y,\epsilon)$. Note that the increase observed 
in the position of this peak is essentially the counterpart of the 
decay of $P(0,\epsilon)$ as a function of $\epsilon$. 
We see here two distinct power-law behaviors, separated by a 
characteristic shear strain $\epsilon_{xy}\approx 0.75 \%$. 
This characteristic shear strain is of the same order of magnitude
as the shear strain $\epsilon_{xy}^{(p)}$ separating the linear
behavior and the plastic flow (see section 1). It is interesting
to note that this characteristic shear strain appears as well when
the linear part of the shear-stress relation is not considered,
that is in the pure plastic flow, while the same figure,
restricted to the linear part of the stress-strain relation, gives
only the first power-law decay. This means that the plastic flow
contains a succession of elastic and plastic events with a small
strain behavior different from the large strain behavior, in
agreement with previous results obtained on the mechanical study
of flowing foams ~\cite{Aubouy}.

The {\it hyper-diffusive} motion of the first bump shown in 
figure~\ref{figfuru}-b, increasing as $\epsilon^{1.4}$, 
implies hyper-diffusive motion of the particles at small shear strain. 
It is in perfect agreement with the variation
$P(0,\epsilon)\propto \epsilon^{-1.4}$ for small
$\epsilon$  and  is related to a strong non-gaussian
behavior. This non-gaussian behavior at small imposed strain has
already been observed in other amorphous systems like foams
~\cite{Diff4}, or granular materials ~\cite{Diff5,Diff6}. In these
systems however, the corresponding exponents, as measured in
~\cite{Diff6} for example, can be different. The exponent
characterizing the hyper-diffusive motion of the particles dragged
by the vortex motion in the linear regime seems thus to be
material dependent, while the non-gaussian character of the motion
could be a characteristic of the small strain deformation at zero
temperature, in disordered systems.

Finally, we have seen here that, even at zero temperature, the
disorder inherent to amorphous systems is sufficient to create,
under sufficiently large external sollicitations with a marked
dissipative behavior (i.e. in the plastic flow regime), a
diffusive motion for the individual particles. Further
investigation of this diffusion process and other nonequilibrium
transport processes -e.g. mobility under an external force- could
allow us to explore the idea of effective temperature in these
systems ~\cite{Diff4,Diff7}.

\section{Conclusion}
\label{conclusion}

 We have shown that the quasi-static dissipative behavior of a two 
dimensional model glass is due to two different kinds of
microscopic events. First, mainly quadrupolar local
rearrangements involving only very few particles  are present in
the linear part of the stress-strain relation. These events
involve only small energy release. Second, in the plastic flow
regime (plateau of the stress strain curve), the plasticity is
dominated by large scale events that involve rearrangements along
lines parallel to the average shear direction. These events have a
very broad distribution of energy losses, which   overlaps with
the one  associated with the more localized events. These
elementary shear bands are correlated in space over distances that
are typical of the elastic inhomogeneity in the medium, and appear
to propagate randomly throughout the system. We suggested that
this behavior could be a source of shear localization in
extended systems. Note that the boundary conditions are crucial
to determine the orientation of the elementary shear bands, 
as already pointed out in ~\cite{maloney}; and they affect also the 
random propagation of shear bands in the system. In our case, the
elementary shear bands reach very quickly the boundaries of our system, 
giving rise to confined motion, and a possible memory free behavior 
of the elementary shear bands was not observable for systems of this size.   
We must finally insist on the differences between the two kinds of dissipative 
events we have clearly pointed out here: particularly the spatial distribution
of localized events is far from the one-dimensional alignement shown in the
elementary shear bands. This is one of the crucial points of this paper to 
distinguish between the two.

We also analyzed the motion of individual particles, 
driven by the plastic deformation.
We only consider the motion in the direction perpendicular to the
average shear velocity. This motion also is, even at zero temperature,
a two step process. In a first step, the motion is
hyper-diffusive and non-gaussian; such displacements are the only
ones that are observed in the linear part of the stress strain
curve. Purely diffusive behavior is however observed on larger length
scales, of the order of a particle size. This diffusive motion
dominates  the plastic flow regime, for sufficiently large imposed
strain. Further exploration of the transport properties within
this plastic flow regime will be the object of future work.

\begin{acknowledgments}
During the course of this work, we had valuable discussions
from L.~Bocquet, M.L.~Falk, A.~Lema\^itre, C.~Maloney, S.~Roux, and D.~Vandembroucq.
Computational support by IDRIS/France, CINES/France and CEA/France is acknowledged.
\end{acknowledgments}

%


\newpage

\newpage
\begin{figure}[t]
\rsfig{FIG-PLAST/FIG-ART/HistoDsxy100pts.eps}
\rsfig{FIG-PLAST/FIG-ART/nDsxy-L-norm.eps}
\vspace*{0.8cm}
\caption[]{(Color online)
(a) Distribution of stress drops in the linear regime (line without symbols, averaged over 10 configurations), and in the regime of plastic flow (one configuration with 34000 steps). In the linear regime, only small size events are present. The various curves correspond to different values of the maximum $u_{max}$ of the non-affine displacements. The distributions are not normalized: the total number of stress drops is shown, in order to compare the contribution of events of various amplitude.

(b) Distribution of stress drops for different system sizes. The value of the stress drops is multiplied by the lateral size $L_y$ in order to show the $1/L_y$ behaviour discussed in the text.

\label{figdrops}}
\end{figure}

\newpage
\begin{figure}[t]
\rsfig{FIG-PLAST/FIG-ART/VM-Invert-1-2339.eps}
\vspace*{0.8cm}
\caption[]{(Color online)
Cumulative displacement in the x-direction (direction of shearing), at various distances y from the shearing wall, and for a total applied strain $\epsilon_{xy}\approx 15\%$. A large shear band appears here in the center.
\label{figshearbande}}
\end{figure}

\newpage

\newpage

\newpage
\begin{figure}[t]
\rsfig{FIG-PLAST/FIG-ART/FigQuadru.eps}
\vspace*{0.8cm}
\caption[]{(Color online)
Magnitude of the quadrupolar projections of the radial ($A_r$) and orthoradial ($A_\theta$) components of the non-affine displacement during a local plastic rearrangement in the linear part of the stress-strain relation. Periodic boundary conditions are responsible for the bump shown at large distances.
\label{figproj}}
\end{figure}

\newpage

\newpage
\begin{figure}[t]
\rsfig{FIG-PLAST/FIG-ART/VMoy-Invert-2338-2339.eps}
\rsfig{FIG-PLAST/FIG-ART/y-bande-nonaff2.eps}
\vspace*{0.8cm}
\caption[]{(Color online)
(a) Averaged horizontal displacement associated to a single but large event. The y position, where the non-affine displacement field is equal to zero, allows to determine the center of the elementary shear band.

(b) Position $y_B$ of the centers of elementary shear bands (determined as described previously), here for the 5000 first steps. No localization appears, even for very large deformation ($170\%$ - not shown here).

\label{figbande-ab}}
\end{figure}

\newpage

\newpage
\begin{figure}[t]
\rsfig{FIG-PLAST/FIG-ART/nDyc-bande3.eps}
\rsfig{FIG-PLAST/FIG-ART/nDt-bande4.eps}
\vspace*{0.8cm}
\caption[]{(Color online)
(a) Distribution of the distances $\delta y_B$ between successive elementary shear bands. The characteristic distance of the exponential fit is $\xi \approx 30 \sigma$.

(b) Distribution of the distances $\delta l_{wall}$ covered by the shearing wall, between successive elementary shear bands. The distribution is exponential, with a characteristic distance covered by the upper wall $l_c = 0.13 \sigma$ (that corresponds to $13$ increments of the strain imposed at a wall, or $\epsilon_{xy}=0.065 \%$).
\label{figbande-cd}}
\end{figure}

\newpage

\newpage

\end{document}